\documentclass[10pt, twocolumn]{IEEEtran}
\usepackage{epsfig,latexsym}
\usepackage{float}
\usepackage{indentfirst}
\usepackage{amsmath}
\usepackage{amssymb}
\usepackage{times}
\usepackage{subfigure}
\usepackage{psfrag}
\usepackage{cite}
\usepackage{lastpage}
\usepackage{fancyhdr}
\usepackage{color}
 \usepackage{amsthm}
\usepackage{bigints}
\begin{document}
\title{ {\huge   MIMO-NOMA Design for Small Packet Transmission in the   Internet of Things }}

\author{ Zhiguo Ding, \IEEEmembership{Senior Member, IEEE}, Linglong Dai, \IEEEmembership{Senior Member, IEEE}, and  H. Vincent Poor, \IEEEmembership{Fellow, IEEE}\thanks{
Z. Ding and H. V. Poor   are with the Department of
Electrical Engineering, Princeton University, Princeton, NJ 08544,
USA (email: $\{$z.ding, poor$\}@$princeton.edu).   Z. Ding is also with the School of
Computing and Communications, Lancaster
University, LA1 4WA, UK (email: z.ding@lancaster.ac.uk). L. Dai is  with the Tsinghua National Laboratory for Information
Science and Technology, Department of Electronic Engineering, Tsinghua
University, Beijing 100084, China (email: daill@tsinghua.edu.cn).
}\vspace{-2em}} \maketitle
\begin{abstract}
A feature of the Internet of Things (IoT) is that some users in the system need to be served quickly for small packet transmission.  To address this requirement,  a new multiple-input multiple-output non-orthogonal multiple access (MIMO-NOMA) scheme is designed in this paper, where one user is served with its quality of service (QoS) requirement strictly met, and the other user is served opportunistically  by using the NOMA concept. The novelty  of this new scheme is that it  confronts  the challenge  that most existing MIMO-NOMA schemes rely on the assumption that users' channel conditions are different, a strong assumption which may not be valid in practice. The developed precoding and detection strategies  can effectively create a significant difference between the users' effective channel gains, and therefore the potential of NOMA can be realized even if the   users' original channel conditions are similar.  Analytical and numerical results  are provided to demonstrate the performance of the proposed MIMO-NOMA scheme.
\end{abstract}\vspace{-1em}
\section{Introduction}
The fifth generation (5G) of mobile communications has been envisioned  to enable  the future Internet of Things (IoT); however, supporting the IoT functionality in 5G networks is challenging since connecting  billions of smart IoT devices with diversified quality of service (QoS) requirements is not a trivial  task, given the constraint of    scarce bandwidth \cite{NGMN}. Non-orthogonal multiple access (NOMA) provides an ideal solution to provide massive connectivity by efficiently  using the available bandwidth resources \cite{docom}, and has consequently been   included in 3GPP long term evolution (LTE) \cite{3gpp1}.

The key idea of NOMA is to ask the users to share the same   resources, such as frequency channels, time slots, and spreading codes, whereas the power domain is used for multiple access. The performance of NOMA in scenarios with single-antenna  users has been studied in \cite{6692652} and \cite{Nomading}. Achieving user fairness with different channel state information (CSI) in NOMA systems has been addressed in \cite{Krikidisnoma}, and the impact of user pairing on NOMA has been investigated in \cite{Zhiguo_CRconoma}.

Since the use of multiple-input multiple-output (MIMO) techniques brings an extra dimension for further performance improvements, the study of the combination of MIMO and NOMA has received considerable   attention recently.  In \cite{7015589} and \cite{Zhiguo_faiconoma}, the scenario in which users have a single antenna has been considered, and various   algorithmic frameworks for optimizing the design of beamforming   in the NOMA transmission system have been proposed. The sum rate has been used  as an objective function in \cite{7095538} and \cite{7383326}  to formulate various optimization problems in   MIMO-NOMA scenarios.   In \cite{Zhiguo_mimoconoma} a zero-forcing based MIMO-NOMA transmission scheme was proposed without requiring the full CSI at the transmitter. A signal alignment based precoding scheme was developed in \cite{Dingicc16}, and it requires fewer  antennas at the users compared to the scheme proposed in \cite{Zhiguo_mimoconoma}. A more detailed literature review can be found in \cite{7263349}.

In this paper, we consider a MIMO-NOMA downlink transmission scenario with one transmitter sending data to two users, e.g., an access point is serving two IoT devices. The feature of the IoT that   users    have diversified QoS requirements is utilized for  the design of the MIMO-NOMA transmission. Particularly, we consider that user $1$ needs to be served quickly  for small packet transmission, i.e., with a low targeted data rate, and user $2$ is to be served with the best effort. Take intelligent transportation as an example, where user $1$ can be a vehicle  receiving the  incident warning information which is contained within a few bytes only. User $2$ can be another vehicle which is to perform some background tasks, such as downloading   multimedia files.   The use of NOMA prevents the drawback of conventional orthogonal multiple access (OMA) that user $1$ whose targeted data rate is small  is served with a dedicated orthogonal channel use. With NOMA, the users are served at the same time/frequency/code, which means that  the bandwidth resources which are solely occupied by user $1$ in the case of OMA can be released to user $2$ in NOMA.

 Most existing NOMA schemes rely on a key assumption that the users' channel conditions are very different. Take the MIMO-NOMA schemes proposed in \cite{7095538} and \cite{Zhiguo_mimoconoma} as   examples. Within the NOMA user pair, one user is assumed to be deployed  close to the base station, and the other is far away from the base station. As shown in  \cite{Zhiguo_CRconoma}, this difference in users' channel conditions is crucial  for realizing the potential of NOMA. However, in practice,  it is very likely that the users who want to participate in NOMA have similar channel conditions. Take our considered IoT scenario as an example, where the two users are categorized by their QoS requirements, not by their channel conditions. It is important to point out that the situation in which users have similar channel conditions  can make the benefits of implementing many existing NOMA schemes very marginal.

 The main contribution of this paper is to  design two sets of  system parameters, precoding and power allocation coefficients, in order to ensure   that the potential of NOMA can be realized even if the users' channel conditions are similar. {\it Firstly}, the precoding matrix at the base station is designed to degrade the effective channel gains at user $1$ and improve the effective channel gains at user $2$, at the same time. As a result, the users' effective channel conditions become very different, an ideal situation for the application of NOMA. The reason to degrade user $1$'s channel condition is that this user is regarded as an IoT user to be served with a low data rate, and therefore, a weaker channel condition could still accommodate this low data rate.  {\it Secondly}, the power allocation coefficients are carefully designed to ensure that user $1$'s QoS requirements can be still met with its degraded channel conditions. Two types of power allocation policies are developed in this paper. One is to meet the user's QoS requirements in the long term, e.g., its targeted outage probability can be satisfied. The other is to realize the user's QoS requirements instantaneously, e.g., the power allocation coefficients are designed to realize user $1$'s  targeted data rate for each channel realization.

 Analytical results are developed to better demonstrate the performance of the proposed MIMO-NOMA scheme. With the long term power allocation policy, the developed analytical results show that user $1$'s targeted outage probability can be strictly guaranteed, and the diversity gains at user $2$ are the same as in the case when user $2$ is served alone. With the short term power allocation policy, user $1$'s outage experience is the same as in the case when all the power is given to user $1$, and the diversity gain achieved at user $2$ is always one. Therefore, between the two power allocation polices, user $2$ prefers   the long term one since the diversity gain it can obtain is larger. However, the short term power allocation policy can ensure that user $1$'s QoS requirement is met instantaneously, a property important to those safety-critical and real-time applications in the IoT.

\section{System Model}\label{section system model}
Consider a MIMO-NOMA downlink transmission scenario with one base station and two users, where the base station is equipped with $M$ antennas and each user is equipped with $N$ antennas. The $N\times M$ channel matrices of two users are denoted by $\mathbf{H}_1$ and $\mathbf{H}_2$, respectively. Elements of the channel matrices    are identically and independent complex Gaussian distributed with zero mean and unit variance. In this paper,  we focus on the scenario without path loss, i.e., two users have similar channel conditions, which is a challenging situation  to realize the potential of NOMA.
Furthermore, we assume $M\geq N$, a scenario in which   existing MINO-NOMA schemes, such as the ones   in \cite{Zhiguo_mimoconoma} and \cite{Dingicc16}, cannot work properly.

Without loss of generality, we assume that user $1$ needs to be connected quickly   to transmit small packets. For example, this user can be an IoT device that    needs to be served with a small predefined  data rate.
The base station will transmit the following vector:
\begin{align}\label{system 2}
\mathbf{x} = \mathbf{P}\mathbf{s},
\end{align}
where   $\mathbf{P}$ is an $M\times N$ precoding matrix. The information bearing vector $\mathbf{s}$ is constructed  by using the NOMA approach as follows: \begin{align}\label{vector x}
\mathbf{s}=\begin{bmatrix} \alpha_1 s_{1}+ \beta_1 w_{1} &\cdots &\alpha_N s_{N}+ \beta_{N}w_{N}\end{bmatrix}^T,
\end{align}
where   $s_i$ is the $i$-th stream transmitted to user $1$, $\alpha_i$ is the power allocation coefficient for $s_i$,   $w_i$ and $\beta_i$ are defined similarly, and $\alpha_k^2+\beta_{k}^2=1$.

As can be seen from \eqref{system 2} and \eqref{vector x}, there are two sets of parameters to be designed,  the precoding matrix $\mathbf{P}$ and the power allocation coefficients $\alpha_i$ ($\beta_i$).  The aim of the proposed design is to realize two goals simultaneously. One is to meet the QoS requirement at user $1$ strictly and the other is to improve  user 2's experience in an opportunistic manner. Alternatively, one can view the addressed NOMA scenario as a special case of cognitive ratio networks, where user $1$ is a primary user whose QoS requirements need to be satisfied strictly and user $2$ is served opportunistically \cite{Zhiguo_CRconoma}.

Assume  that the QR decomposition of user $2$'s channel matrix, $\mathbf{H}_2$, is given by
\begin{align}
\mathbf{H}_2^H=\mathbf{Q}_2\tilde{\mathbf{R}}_2,
\end{align}
where $\mathbf{Q}_2$ is an $M\times M$ unitary matrix, and $\tilde{\mathbf{R}}_2$ is an $M\times N$ matrix obtained from the QR decomposition. Define $\mathbf{V}_2$ as an $M\times N$ matrix collecting the $N$ left columns of $\mathbf{Q}_2$, and $\mathbf{R}_2$ is an $N\times N$ upper submatrix of $\tilde{\mathbf{R}}_2$. From the QR decomposition, we know that $\mathbf{H}_2^H=\mathbf{V}_2\mathbf{R}_2$.
 The precoding matrix $\mathbf{P}$ is set as $\mathbf{P}=\mathbf{V}_2$, which   is to improve the signal strength at user $2$. As can be seen from the following subsection,   this choice of the precoding matrix also  degrades the channel conditions at user $1$, which makes user $1$  analogous  to a cell edge user  in a conventional NOMA setup.

User $2$'s observation can be expressed as follows:
\begin{align}
\mathbf{y}_2  =\mathbf{R}_2^H\mathbf{s} +\mathbf{n}_2,
\end{align}
where $\mathbf{n}_2$ is the noise vector. Since $\mathbf{R}_2^H$ is a lower triangular  matrix,
successive interference cancellation (SIC) can be carried out to cancel inter-layer interference (between $w_i$ and $w_j$, $i\neq j$) and intra-layer interference (between $s_i$ and $w_i$). Particularly, suppose that $s_j$ and $w_j$ from the previous  layers, $j<i$, are decoded successfully, whose outage probability will be included for the calculation of the overall  probability in the next section. User $2$ can decode the message intended for  user $1$ at the $i$-th layer, $s_i$, with the following signal-to-interference-plus-noise ratio (SINR):
\begin{align}\label{SINR}
\text{SINR}_{2,i'} = \frac{\alpha_i^2 [\mathbf{R}_2^H]^2_{i,i}}{\beta_i^2[\mathbf{R}_2^H]^2_{i,i}+\frac{1}{\rho} },
\end{align}
where  $\rho$ denotes the transmit signal-to-noise ratio (SNR) and $[\mathbf{A}]_{i,j}$ denotes the element at the $i$-th row and the $j$-th column of $\mathbf{A}$. Denote the targeted data rate of user $m$ at the $i$-th layer by $R_{m,i}$.
Provided that $\log(1+\text{SINR}_{2,i'})>R_{1,i}$, user $2$ can   successfully remove user $1$'s message, $s_i$, from its $i$-th layer, and its own message can be decoded with the following SNR:
\begin{align}\label{SNR}
\text{SNR}_{2,i} =\rho \beta_i^2 [\mathbf{R}_2^H]^2_{i,i}.
\end{align}

 User $1$'s observation    is given by
\begin{align}\label{system 1}
\mathbf{y}_{ 1} = \mathbf{H}_{1}\mathbf{P}\mathbf{s}+\mathbf{n}_{ 1},
\end{align}
where  $\mathbf{n}_{ 1}$ is an $N\times 1$ noise vector. Analogously  to the cell edge user in a conventional NOMA network,  user $1$ is not to decode $w_i$, which means that the use of the QR based detection will result in significant performance loss, as discussed in Section \ref{QR subsection}.   Therefore, zero forcing is applied at user $1$. Particularly, the system model at user $1$ can be written as follows:
\begin{align}\label{system 3}
\left(\mathbf{H}_{1}\mathbf{V}_2\right)^\dagger \mathbf{y}_{ 1} &=\mathbf{s}+\left(\mathbf{H}_{1}\mathbf{V}_2\right)^\dagger \mathbf{n}_{ 1},
\end{align}
where $\left(\mathbf{H}_{1}\mathbf{V}_2\right)^\dagger = \left(\mathbf{V}_2^H \mathbf{H}_{1}^H\mathbf{H}_{1}\mathbf{V}_2\right)^{-1}\mathbf{V}_2^H \mathbf{H}_{1}^H$. It is worth pointing out that the dimension of $\mathbf{V}_2$ is $M\times N$, and therefore $\mathbf{H}_1\mathbf{V}_2$ is an $N\times N$ square matrix, which means $\left(\mathbf{H}_{1}\mathbf{V}_2\right)^\dagger=\left(\mathbf{H}_{1}\mathbf{V}_2\right)^{-1}$. It is assumed here that the channel matrices are full column rank.  As a result, user 1 can decode its  message at the $i$-th layer  with the following SINR:
\begin{align}
\text{SINR}_{1,i} = \frac{ \alpha_i^2 z_i }{ \beta_i^2z_i+\frac{1}{\rho}},
\end{align}
where $z_i=\frac{1}{\left[\left(\mathbf{V}_2^H \mathbf{H}_{1}^H\mathbf{H}_{1}\mathbf{V}_2\right)^{-1}\right]_{i,i}}$.

\subsection{Impact of the Proposed Precoding Scheme}
The two users' experiences with the proposed precoding scheme are different.
 According to the previous discussions, the reception reliability at user $2$ is determined by the parameter $x_i$, where $x_i\triangleq[\mathbf{R}_2^H]^2_{i,i}$.  Denote an $M\times (M-N)$ complex Gaussian matrix which is independent of $\mathbf{H}_2$ by $\mathbf{B}$. The QR decomposition of $\begin{bmatrix} \mathbf{H}_2^H & \mathbf{B} \end{bmatrix}$ is given by
\begin{align}
\begin{bmatrix} \mathbf{H}_2^H & \mathbf{B} \end{bmatrix}=\mathbf{Q}_2\bar{\mathbf{R}}_2 =\begin{bmatrix} \mathbf{V}_2 &\bar{\mathbf{V}}_2\end{bmatrix} \underset{\bar{\mathbf{R}}_2}{\underbrace{\begin{bmatrix}\mathbf{R}_2 &\mathbf{C} \\ \mathbf{0}_{(M-N)\times N} & \mathbf{D}\end{bmatrix}}},
\end{align}
where $\bar{\mathbf{R}}$ is an $M\times M$ upper triangular matrix, $\bar{\mathbf{V}}_2$ is a submatrix of $\mathbf{Q}_2$,   $\mathbf{C}$ and $\mathbf{D}$ are the submatrices of $\bar{\mathbf{R}}_2$.   According to \cite{Tuliobook}, the elements of $\bar{\mathbf{R}}$ are independent, and the square of the $i$-th element on its diagonal follows the chi-square distribution with $2(M-i+1)$ degrees of freedom, i.e.,  $f_{x_i}(x) = \frac{x^{M-i}}{(M-i)!}e^{-x}$. Therefore,   more antennas at the base station can improve the receive signal strength at user $2$ which is a function of $[\mathbf{R}_2^H]^2_{i,i}$.

On the other hand, the reception reliability at user $1$ is degraded  due to the use of the precoding matrix, $\mathbf{P}$, as explained in the following.  Note that $\mathbf{V}_2$ is a unitary matrix obtained from the QR decomposition based on $\mathbf{H}_2$. Because $\mathbf{H}_1$ and $\mathbf{H}_2$ are independent, and also by using the fact that a unitary transformation of a Gaussian matrix does not alter its statistical properties, $\mathbf{H}_{1}\mathbf{P}$ is still an $N\times N$ complex Gaussian  matrix, which means that the use of the proposed precoding matrix {\it shrinks} user $1$'s channel matrix from an $M\times N$ complex Gaussian matrix to another  complex Gaussian  matrix  with smaller size. Note that  the probability density function (pdf) of the effective  channel gain, $\frac{1}{\left[\left(\mathbf{V}_2^H \mathbf{H}_{1}^H\mathbf{H}_{1}\mathbf{V}_2\right)^{-1}\right]_{i,i}}$, is given by
 \begin{align}\label{pdf}
f_{ \frac{1}{\left[\left(\mathbf{V}_2^H \mathbf{H}_{1}^H\mathbf{H}_{1}\mathbf{V}_2\right)^{-1}\right]_{i,i}}}(z) =e^{-z},
 \end{align}
 which is no longer a function of $M$.

The impact of the proposed precoding scheme can be clearly illustrated by using the following extreme example. Consider a special case with $N=1$, where the channel matrices become $1\times M$ vectors, denoted by $\mathbf{h}_1$ and $\mathbf{h}_2$, respectively. After applying $\mathbf{P}$,  the effective channel gain at user $2$ is $|\mathbf{h}_2|^2$ which becomes stronger by increasing $M$. On the other hand, the effective channel gain at user $1$ is always exponentially distributed, and the use of more antennas at the base station does not improve the transmission reliability at user $1$.
\subsection{Power Allocation Policies}
Because the precoding matrix degrades  user $1$'s channel conditions, the power allocation coefficients  $\alpha_i$  ($\beta_i$) needs to be carefully designed to ensure that user $1$'s QoS requirements are met,  which motivates the following two power allocation policies.
\subsubsection{Power allocation policy I} This approach  is to meet user $1$'s  QoS requirements in the long term.
Recall that the targeted data rate for user $1$ to decode its  message at the $i$-th layer ($s_i$) is denoted by $R_{1,i}$. As a result, the outage probability for user $1$ to decode $s_i$  is given by
\begin{align}
\mathrm{P}_{1,i}^o\triangleq \mathrm{P}\left(\log(1+\text{SINR}_{1,i})< R_{1,i}\right).
\end{align}
The power allocation coefficients, $\alpha_i$ and $\beta_i$, are designed  to satisfy  the following constraint:
\begin{align}\label{policy I}
\mathrm{P}_{1,i}^o\leq \mathrm{P}_{1,i,\text{target}},
\end{align}
where $\mathrm{P}_{1,i,\text{target}}$ denotes the targeted outage probability. A closed-form expression for the power allocation coefficients is given in Section \ref{subsection p1 u1}.  The advantage of this type of power allocation is that there is no need to update power allocation coefficients frequently, but it cannot satisfy user $1$'s QoS requirements instantaneously.
\subsubsection{Power allocation policy II} This approach  is to meet user $1$'s QoS requirements instantaneously.
 This type of power allocation is quite similar to the cognitive radio inspired power allocation policy proposed in~\cite{Zhiguo_CRconoma}. Particularly, the power allocation coefficients are defined to ensure that the targeted data rate of user $1$ is met instantaneously, i.e.,
\begin{align}\label{policy II}
\log(1+\text{SINR}_{1,i})\geq R_{1,i}.
\end{align}
By defining $\epsilon_{k,i}=2^{R_{k,i}}-1$, the above constraint yields  the following power allocation policy:
\begin{align}\label{beta choice}
\beta_{i}^2 = \max\left\{0, \frac{z_i-\frac{\epsilon_{1,i}}{\rho}}{z_i(1+\epsilon_{1,i})}\right\}.
\end{align}
The above policy is sufficient for those scenarios addressed in   \cite{Zhiguo_CRconoma} and \cite{Zhiguo_mimoconoma}, where  users are  {\it ordered} according to their channel conditions. For the scenario addressed in this paper,     $z_i< x_i$ does not always hold, and it is possible that the effective channel gain of user $1$ is stronger. If $z_i>x_i$,  the value of $\beta_i$ in \eqref{beta choice} is a very poor  choice for  user $2$,
as it is guaranteed that SIC at user 2 will fail. To ensure that SIC at user $2$ can be carried out successfully, we revise the power allocation strategy as follows:
 {\small \begin{align}\label{beta choice new}
\beta_{i}^2 =  \min\left\{\max\left\{0,\frac{z_i-\frac{\epsilon_{1,i}}{\rho}}{z_i(1+\epsilon_{1,i})}\right\},
\max\left\{0,\frac{x_i-\frac{\epsilon_{1,i}}{\rho}}{x_i(1+\epsilon_{1,i})}, \right\}\right\}.
\end{align}}
This is to ensure that user $1$'s message can be decoded by  both users with the best effort.
 In an extreme case with $x_i\rightarrow 0$ and a fixed $z_i$, $\beta_i=0$, i.e., the cognitive
radio user, user $2$, will not be admitted.

Note that it is also possible to reverse  the decoding order when $z_i>x_i$. In this case, we need to ensure that the following two conditions are satisfied. One is to ensure that user $1$ can decode the message intended for user $2$, $w_i$,
\begin{align}
\log\left(1+ \frac{z_i\beta_i^2}{z_i\alpha_i+\frac{1}{\rho}}\right)\geq R_{2,i},
\end{align}
and the other is to ensure user $1$ can decode its own message,
\begin{align}
\log\left(1+ \rho z_i\alpha_i^2\right)\geq R_{1,i}.
\end{align}
As a result,   user $1$'s outage experience becomes  a function of $R_{2,i}$ which is user $2$'s targeted data rate, since user $1$ needs to decode user $2$'s message first. When user $2$'s targeted data  rate is varying, there is an uncertainty as to whether
user $1$'s QoS requirements can be met strictly. Therefore, this type of  reverse  NOMA decoding is not considered in this paper.

  \section{Outage Performance at User $1$}
  User $1$'s outage performance will be studied in the following two subsections with the two different power allocation policies.
  \subsection{Power Allocation Policy I}\label{subsection p1 u1}
 Recall that the outage probability for user $1$ to detect $s_i$ can be expressed as follows:
 \begin{align}
 \mathrm{P}_{1,i}^0 &=   \mathrm{P}\left(
 x_i<\frac{\frac{\epsilon_{1,i}}{\rho}}{\alpha_i^2-\beta_i^2\epsilon_{1,i}}\right),
 \end{align}
When power allocation policy I is adopted,  the power allocation coefficients are set to meet user $1$'s QoS requirements  in the long term, which means that neither $\alpha_i$ or $\beta_i$ is a function of instantaneous channel gains. Therefore,  the outage probability in this case is the cumulative distribution function (CDF) of $
 \frac{1}{\left[\left(\mathbf{V}_2^H \mathbf{H}_{1}^H\mathbf{H}_{1}\mathbf{V}_2\right)^{-1}\right]_{i,i}}$.

By using the pdf in \eqref{pdf}, the outage probability can be expressed as follows:
 \begin{align}\label{ana1}
 \mathrm{P}_{1,i}^0 &=   1-e^{-\frac{\frac{\epsilon_{1,i}}{\rho}}{\alpha_i^2-\beta_i^2\epsilon_{1,i}}}.
 \end{align}
 In order to ensure $\mathrm{P}_{1,i}^o\leq \mathrm{P}_{1,i,\text{target}}$, the power allocation coefficients need to be set as follows:
 \begin{align}\label{betax1}
 \beta^2_i = \frac{1+\frac{\epsilon_{1,i}}{\rho\ln (1-\mathrm{P}_{1,i,\text{target}})}}{1+\epsilon_{1,i}}.
 \end{align}
 Note that $1-\mathrm{P}_{1,i,\text{target}}\leq 1$, which means $\ln (1-\mathrm{P}_{1,i,\text{target}})\leq 0$. Therefore, the choice of $\beta_i$ in \eqref{betax1} is always smaller than or equal to one, i.e., $\beta_i\leq  1$. In order to ensure the choice of $\beta_i$ in \eqref{betax1} positive or equivalently  $\frac{1+\frac{\epsilon_{1,i}}{\rho\ln (1-\mathrm{P}_{1,i,\text{target}})}}{1+\epsilon_{1,i}}> 0$, the following constraint is imposed on the targeted outage probability:
 \begin{align}\label{range1}
 1>\mathrm{P}_{1,i,\text{target}}> 1-e^{-\frac{\epsilon_{1,i}}{\rho}}.
 \end{align}
 We ignore the choice of $\mathrm{P}_{1,i,\text{target}}=1$, since this choice does not consider user $1$'s QoS requirements.
 The righthand  side of the above equation  is a lower bound of the targeted outage probability which is achieved by giving all the power to user $1$. Or in other words, if the targeted outage probability is smaller than or equal to $\left(1-e^{-\frac{\epsilon_{1,i}}{\rho}}\right)$, we will have $\beta_i=0$ and the addressed NOMA scenario is degraded to the case in which  only user $1$ is served. Therefore, in the remainder  of this paper,  it is assumed that the targeted outage probability is chosen to be larger than $\left(1-e^{-\frac{\epsilon_{1,i}}{\rho}}\right)$ when power allocation policy I is used, in order to avoid the trivial case of  $\beta_i=0$.
\subsection{Power Allocation Policy II}
While the power allocation coefficients are set to ensure $\log(1+\text{SINR}_{1,i})\geq R_{1,i}$, outage   can still occur at user $1$  since this ideal choice of power allocation  might not be feasible  due to deep fading, i.e., a situation with very small channel gains can result in  $\beta_i=0$.
 Rewrite the expression of $\beta_i$ in \eqref{beta choice new} as follows:
  \begin{align}\label{beta choice x}
\beta_{i}^2 =   \min\left\{\beta_{i,z}^2, \beta_{i,x}^2
 \right\} ,
\end{align}
where $\beta_{i,z}^2=\max\left\{0, \frac{z_i-\frac{\epsilon_{1,i}}{\rho}}{z_i(1+\epsilon_{1,i})}\right\}$,  $\beta_{i,x}^2=\max\left\{0,
\frac{x_i-\frac{\epsilon_{1,i}}{\rho}}{x_i(1+\epsilon_{1,i})}\right\}$, $\alpha_{i,x}$ and $\alpha_{i,z}$ are defined similarly.
 Note that when $x_i>z_i$, $\beta_{i,x}\geq\beta_{i,z}$, otherwise  $\beta_{i,x}\leq \beta_{i,z}$.
With these definitions, the outage probability for user $1$ to detect $s_i$ can be expressed as follows:
 \begin{align}
 \mathrm{P}_{1,i}^0 &= \mathrm{P}\left(\log(1+\text{SINR}_{1,i})<R_{1,i}\right) \\ \nonumber &= \underset{T_1}{\underbrace{\mathrm{P}\left(x_i>z_i, \log\left(1+\frac{z_i \alpha_{i,z}^2}
 {z_i\beta_{i,z}^2+\frac{1}{\rho}}\right)<R_{1,i}\right)}}\\ \nonumber &+\underset{T_2}{\underbrace{\mathrm{P}\left(z_i>x_i, \log\left(1+\frac{z_i \alpha_{i,x}^2}
 {z_i\beta_{i,x}^2+\frac{1}{\rho}}\right)<R_{1,i}\right)}}.
 \end{align}
First consider the case of $x_i>z_i$. If $\beta_{i,z}\neq 0$, we have $\log\left(1+\frac{z_i \alpha_{i,z}^2}
 {z_i\beta_{i,z}^2+\frac{1}{\rho}}\right)=R_{1,i}$, which means that no outage occurs. Therefore, the outage event   when $x_i>z_i$ is due to $\beta_i=0$, and therefore, $T_1$ can be simplified  as follows:
\begin{align}
T_1 &=\mathrm{P}\left(x_i>z_i, z_i<\frac{\epsilon_{1,1}}{\rho}\right).
\end{align}

The second factor $T_2$ can be expressed as follows:
\begin{align}
T_2 &=\mathrm{P}\left(x_i<z_i<\frac{\frac{\epsilon_{1,i}}{\rho}}{\alpha_{i,x}^2-\beta_{i,x}^2\epsilon_{1,i}}\right)\\\nonumber &=\mathrm{P}\left(x_i<z_i<\frac{\frac{\epsilon_{1,i}}{\rho}}{1-\max\left\{0,
\frac{x_i-\frac{\epsilon_{1,i}}{\rho}}{x_i(1+\epsilon_{1,i})}\right\}(1+\epsilon_{1,i})}\right).
 \end{align}
In order to explicitly show the outage events, the factor $T_2$ can be written as follows:
\begin{align}
T_2 &=\mathrm{P}\left(x_i<z_i<\frac{\frac{\epsilon_{1,i}}{\rho}}{\min\left\{1,
\frac{\frac{\epsilon_{1,i}}{\rho}}{x_i}\right\}}\right)\\ \nonumber &=\mathrm{P}\left(x_i<z_i<\max\left\{ \frac{\epsilon_{1,i}}{\rho},
x_i\right\}\right)
\\ \nonumber &=\mathrm{P}\left(x_i<z_i<\frac{\epsilon_{1,i}}{\rho} \right) +\mathrm{P}\left(x_i<z_i<
x_i, \frac{\epsilon_{1,i}}{\rho}<
x_i\right).
 \end{align}
 Since the second probability in the above equation is zero, the overall outage probability  can be calculated as follows:
 \begin{align}\label{outage user 1}
\mathrm{P}_{1,i}^0 & = \mathrm{P}\left(x_i>z_i, z_i<\frac{\epsilon_{1,1}}{\rho}\right)+\mathrm{P}\left(x_i<z_i<\frac{\epsilon_{1,i}}{\rho}\right)
 \\ \nonumber &= \mathrm{P}\left( z_i<\frac{\epsilon_{1,1}}{\rho}\right)=1-e^{-\frac{\epsilon_{1,i}}{\rho}},
 \end{align}
 which means that the diversity gain for user $1$ to decode $s_i$ is one.

{\it Remark 1:} Consider a benchmark scheme with $\beta_i=0$, i.e., user $2$ is not served at all. By using   zero forcing detection at user $1$'s receiver, it is straightforward to show that the outage probability achieved by this benchmark scheme is exactly the same as the one in \eqref{outage user 1}. The reason for this phenomenon is that power allocation policy II is to ensure that the QoS requirements of user $1$ is met instantaneously, while user $2$ is served under the condition that the outage probability at user $1$ is not degraded compared to the case with  user $1$ served alone.

 {\it Remark 2:}  Another interesting benchmark scheme is to consider that  the precoding matrix is designed for user $1$ by using the zero forcing approach, i.e., $\mathbf{P}=\left( \mathbf{H}_{1}^H\mathbf{H}_{1}\right)^{-1}\mathbf{H}_{1}^H$. It is straightforward to show that this benchmark scheme achieves a diversity gain of $(M-N+1)$ for each stream, by following   steps similar to those in \cite{Zhiguo_mimoconoma}.  This diversity gain loss is because   the precoding matrix $\mathbf{P}$ proposed in this paper shrinks user $1$'s channel matrix from an $N\times M$ complex Gaussian matrix to an $N\times N $ complex Gaussian matrix.  This degradation  is caused on purpose in order to ensure  that the two users' channel conditions become very different.

 \subsection{When User $1$ Adopts  the QR Based Approach}\label{QR subsection}
 Instead of zero forcing, user $1$ can also use  the QR based approach for detection. In the following, we will show that the performance of the QR based approach is worse than that of the zero forcing   one introduced in the previous section.

 Suppose  that the effective channel matrix at user $1$ has the QR decomposition as $\mathbf{H}_1\mathbf{V}_2=\mathbf{Q}_1\mathbf{R}_1$, and therefore,    the observation at user $1$ can be expressed as follows:
\begin{align}
\mathbf{Q}_1^H\mathbf{y}_1  =\mathbf{R}_1\mathbf{s}+\mathbf{Q}^H_1\mathbf{n}_1.
\end{align}
Recall that $\mathbf{H}_1\mathbf{V}_2$ is an $N\times N$ complex Gaussian matrix. Therefore, $[\mathbf{R}_1]_{i,i}^2$ is chi-squared distributed with $2(N-i+1)$ degrees of freedom. Unlike user $2$, user 1 needs to decode the $i$-th stream first before decoding the $j$-th stream, $N\geq i>j\geq 1$, since $\mathbf{R}_1$ is an upper triangular matrix and $\mathbf{R}_2^H$ is a lower triangular matrix.
Since user $1$ does not need  to decode the messages intended for user $2$,  the system model for  the $i$-th stream at user $1$ can be rewritten as follows:
\begin{align}
\tilde{y}_{1,i} & =[\mathbf{R}_1]_{i,i}\alpha_i s_i + [\mathbf{R}_1]_{i,i}\beta_i w_i\\ \nonumber
&+ \sum^{N}_{j=i+1}\left([\mathbf{R}_1]_{i,j}\alpha_j s_j + [\mathbf{R}_1]_{i,j}\beta_j w_j\right) +n_{1,i}.
\end{align}
where $\tilde{y}_{1,i}$ is the $i$-th element of $\mathbf{Q}_1^H\mathbf{y}_1$ and $n_{1,i}$ is defined similarly. Consider an ideal case in which $s_j$  has been decoded correctly. By using this assumption, the outage probability at user $1$ can be lower bounded as follows:
\begin{align}
\mathrm{P}_{1,i}^o&\geq \mathrm{P}\left(   \frac{[\mathbf{R}_1]^2_{i,i}\alpha_i^2}{[\mathbf{R}_1]_{i,i}^2\beta_i^2 +\sum^{N}_{j=i+1} [\mathbf{R}_1]_{i,j}^2\beta_j^2 +\frac{1}{\rho}}<\epsilon_{1,i}  \right)\\ \nonumber &\geq \mathrm{P}\left( \frac{[\mathbf{R}_1]^2_{i,i}\alpha_i^2}{\sum^{N}_{j=i+1} [\mathbf{R}_1]_{i,j}^2\beta_j^2 +\frac{1}{\rho}} <\epsilon_{1,i} \right),
\end{align}
where $1\leq i<N$.
In order to obtain  some insight, we focus on the case with power allocation policy I, and assume $\beta_j=\beta_i$, for $i\neq j$. Define $u_i=\sum^{N}_{j=i+1} [\mathbf{R}_1]_{i,j}^2$. According to \cite{Tuliobook}, the entries of $\mathbf{R}_1$ are independent, and $[\mathbf{R}_1]_{i,j}^2$ with $i<j$ is exponentially distributed, which means $u_i$ is chi-square distributed with $2(N-i)$ degrees of freedom, i.e., $f_{u_i}(u) = \frac{u^{N-i-1}}{(N-i-1)!}e^{-u}$. It is straightforward to verify that user $1$'s outage probability becomes a non-zero constant, regardless of how large the SNR is. Since $\mathrm{P}_{1,i}^o$ is lower bounded by a non-zero constant, this means that, when the QR based approach is used, the outage probability at user $1$ never goes to zero, even if the transmission power becomes infinite. Recall that  the use of zero forcing can effectively cancel the inter-layer interference at user $1$. For this reason,  only zero forcing detection is considered at user $1$ in this paper.

\section{Outage Performance at User $2$}
Since user $2$ experiences differently with different power allocation policies, two subsections are provided in the following to study the two scenarios.
\subsection{Power Allocation Policy I}\label{subsection p1}
Recall that SIC is carried out at user $2$ to remove both intra-layer and inter-layer interference. The outage event for user $2$ to decode its own message  at the $i$-th layer can be expressed as follows:
\begin{align}\nonumber
\mathcal{O}_{2,i}\triangleq & \underset{m\in \{1, \cdots, i\}}{\bigcup}  \tilde{\mathcal{O}}_{2,m},
\end{align}
where $\tilde{\mathcal{O}}_{2,m} $ denotes an event that user $2$ cannot successfully decode  the messages at the $m$-th layer, $s_m$ and $w_m$, while  all the messages in the previous layers, $s_n$ and $w_n$, for $1\leq n <m$, can be decoded correctly. Note that $ \tilde{\mathcal{O}}_{2,m}\cap  \tilde{\mathcal{O}}_{2,n}=\emptyset$, for $m\neq n$.

Since there are two messages at each layer, the outage event $\tilde{\mathcal{O}}_{2,m} $  can be further expressed as follows:
\begin{align}\nonumber
\tilde{\mathcal{O}}_{2,m}=& \bar{E}_{m,1} \bigcup  \bar{E}_{m,2},
\end{align}
where the two events are defined as follows:
\begin{itemize}
\item $\bar{E}_{m,1}$: the event that user $2$ cannot decode $s_m$, but can decode  all the messages from the previous layers, $s_n$ and $w_n$, for $1\leq n \leq (m-1)$;

\item $\bar{E}_{m,2}$: the event that user $2$ cannot decode $w_m$, but can decode $s_m$, as well as  $s_n$ and $w_n$, for $1\leq n \leq (m-1)$ .
\end{itemize}
Note that $\bar{E}_{m,1}\cap \bar{E}_{m,2}=\emptyset$.

By using the above definitions, the outage probability for user $2$ to decode its own message  at the $i$-th layer can be expressed as follows:
\begin{align}\nonumber
\mathrm{P}_{2,i}^o=& \sum^{i}_{m=1} \left(\mathrm{P}\left( \bar{E}_{m,1} \right)+\mathrm{P}\left( \bar{E}_{m,2} \right)\right).
\end{align}

The first type of the outage probability $\mathrm{P}\left( \bar{E}_{m,1} \right)$ can be expressed as follows:
\begin{align}
\mathrm{P}\left( \bar{E}_{m,1} \right) &= \mathrm{P}\left(\log\left(1+\text{SINR}_{2,m'}\right)<R_{1,m}, \right. \\ \nonumber&
\log\left(1+\text{SINR}_{2,n'}\right)>R_{1,n},\\ \nonumber & \left.\log\left(1+\text{SNR}_{2,n}\right)>R_{2,n}, \forall ~n\in \{1, \cdots, m-1\}\right).
\end{align}
Similarly the outage probability $\mathrm{P}\left( \bar{E}_{m,2} \right)$ can be expressed as follows:
\begin{align}
\mathrm{P}\left( \bar{E}_{m,2} \right) &= \mathrm{P}\left(\log\left(1+\text{SNR}_{2,m}\right)<R_{2,m},\right.\\ \nonumber &\log\left(1+\text{SINR}_{2,m'}\right)>R_{1,m},  \\ \nonumber&
\log\left(1+\text{SINR}_{2,n'}\right)>R_{1,n},\\ \nonumber & \left.\log\left(1+\text{SNR}_{2,n}\right)>R_{2,n}, \forall ~n\in \{1, \cdots, m-1\}\right).
\end{align}
Note that for the case of $m=1$, the above outage probabilities can be simplified as $\mathrm{P}\left( \bar{E}_{1,1} \right) = \mathrm{P}\left(\log\left(1+\text{SINR}_{2,1'}\right)<R_{1,1}\right)$ and $\mathrm{P}\left( \bar{E}_{1,2} \right) = \mathrm{P}\left(\log\left(1+\text{SNR}_{2,1}\right)<R_{2,1}, \log\left(1+\text{SINR}_{2,1'}\right)>R_{1,1}\right).$

By using the SINR expression in \eqref{SINR} and the above definitions, $\mathrm{P}\left( \bar{E}_{m,1} \right)$ can be expressed as follows:
\begin{align}
\mathrm{P}\left( \bar{E}_{m,1} \right) &= \mathrm{P}\left(\log\left(1+\frac{\alpha_m^2 x_m}{\beta_m^2x_m+\frac{1}{\rho} }\right)<R_{1,m}, \right. \\ \nonumber&
\log\left(1+\frac{\alpha_n^2 x_n}{\beta_n^2x_n+\frac{1}{\rho} }\right)>R_{1,n},\\ \nonumber & \left.\log\left(1+\beta_n^2 x_n\right)>R_{2,n}, \forall ~n\in \{1, \cdots, m-1\}\right).
\end{align}
Provided that power allocation policy I is used,  the power coefficients are not a function of instantaneous channel gains, which yields the following:
\begin{align}\label{indepdent}
&\mathrm{P}\left( \bar{E}_{m,1} \right) = \mathrm{P}\left( x_m<  \frac{\frac{\epsilon_{1,m}}{\rho}}{\alpha^2_m - \beta_{m}^2\epsilon_{1,m}}  \right) \\ \nonumber&
\times \prod^{m-1}_{n=1}\mathrm{P}\left( x_n>  \frac{\frac{\epsilon_{1,n}}{\rho}}{\alpha^2_n - \beta_{n}^2\epsilon_{1,n}} , x_n>\frac{\epsilon_{2,n}}{\rho \beta_n^2 } \right).
\end{align}
for $\alpha^2_i > \beta_{i}^2\epsilon_{1,i}$,   $\forall~i\in\{1,\cdots, m\}$, otherwise the probability is one. Note that \eqref{indepdent} follows from the fact that the elements on the diagonal of $\mathbf{R}_2$, $x_m$, are independent. It can be verified that the choice of $\beta_i$ in \eqref{betax1} can always ensure $\alpha^2_i > \beta_{i}^2\epsilon_{1,i}$ since
\begin{align}
\alpha^2_i - \beta_{i}^2\epsilon_{1,i} &=1 - \beta_{i}^2(1+\epsilon_{1,i}) \\\nonumber
&\geq  -\frac{\epsilon_{1,i}}{\rho\ln (1-\mathrm{P}_{1,i,\text{target}})} >0,
 \end{align}
 where $\mathrm{P}_{1,i,\text{target}}<1$ as defined in \eqref{range1}.

By applying the pdf of $x_m$, the above probability can be obtained as follows:
\begin{align}\label{2222}
&\mathrm{P}\left( \bar{E}_{m,1} \right) = \frac{\gamma(M-m+1,\xi_m)}{(M-m)!} \\ \nonumber&
\times \prod^{m-1}_{n=1}\left[1-\frac{\gamma\left(M-n+1,\max\left\{\xi_n, \frac{\epsilon_{2,n}}{\rho \beta_n^2 }\right\} \right)}{(M-n)!}\right] ,
\end{align}
where $\xi_m=\frac{\frac{\epsilon_{1,m}}{\rho}}{\alpha^2_m - \beta_{m}^2\epsilon_{1,m}} $ and $\gamma(\cdot)$ denotes the incomplete gamma function \cite{GRADSHTEYN}.

Similarly the probability of $\mathrm{P}\left( \bar{E}_{m,2} \right)$ can be calculated as follows:
\begin{align}\nonumber
&\mathrm{P}\left( \bar{E}_{m,2} \right) =\prod^{m-1}_{n=1}\left[1-\frac{\gamma\left(M-n+1,\max\left\{\xi_n, \frac{\epsilon_{2,n}}{\rho \beta_n^2 }\right\} \right)}{(M-n)!}\right] \\  &
\times \frac{\left[
\gamma\left(M-m+1,\frac{\epsilon_{2,m}}{\rho \beta_m^2 }\right) -  \gamma(M-m+1,\xi_m)\right]}{(M-m)!},
\end{align}
if $\frac{\epsilon_{2,m}}{\rho \beta_m^2 }\geq \xi_m$, otherwise $\mathrm{P}\left( \bar{E}_{m,2} \right)=0$.

Hence, the outage probability for user $2$ to decode its own message  at the $i$-th layer can be expressed as follows:
\begin{align} \label{policy I user 2}
\mathrm{P}_{2,i}^o=& \sum^{i}_{m=1}
\frac{\gamma\left(M-m+1,\max\{\xi_m, \frac{\epsilon_{2,m}}{\rho \beta_m^2 }\}\right)}{(M-m)!} \\ \nonumber &\times \prod^{m-1}_{n=1}\left[1-\frac{\gamma\left(M-n+1,\max\left\{\xi_n, \frac{\epsilon_{2,n}}{\rho \beta_n^2 }\right\} \right)}{(M-n)!}\right].
\end{align}
At high SNR, i.e., $\rho$ approaches infinity, for a fixed $\mathrm{P}_{1,i,target}$ which is constrained as in \eqref{range1} and not a function of $\rho$, it is straightforward to show that both $\xi_m$ and $\frac{\epsilon_{2,m}}{\rho \beta_m^2 }$ approach zero. Therefore,  the outage probability can be approximated as follows:
\begin{align} \label{policy I user 2 approximation}
\mathrm{P}_{2,i}^o=& \sum^{i}_{m=1}
\left[1-e^{-\gamma_m}\left(\sum^{M-m}_{j=0}\frac{\gamma_m^j}{j!}\right)\right] \\ \nonumber &\times \prod^{m-1}_{n=1}\left[e^{-\gamma_n}\left(\sum^{M-n}_{j=0}\frac{\gamma_n^j}{j!}\right)\right]\\ \nonumber \approx &\sum^{i}_{m=1} \frac{\gamma_m^{M-m+1}}{(M-m+1)!} \approx \frac{\gamma_m^{M-i+1}}{(M-i+1)!} ,
\end{align}
where $\gamma_m=\max\{\xi_m, \frac{\epsilon_{2,m}}{\rho \beta_m^2 }\}$. By using this high SNR approximation, one can ready  find     that the diversity gain for user $2$ to decode $w_i$ is $(M-i+1)$.

{\it Remark 3:} In Section \ref{section simulation}, we will also use another choice of the targeted outage probability, i.e., $\mathrm{P}_{1,i,target}=1-e^{-\frac{x\epsilon_{1,i}}{\rho}}$, where $x$ is not a function of $\rho$ and $x>1$. This targeted outage probability becomes  a function of $\rho$. First note that  this choice of $\mathrm{P}_{1,i,target}$ still  fits the range defined in \eqref{range1}.  Although this choice of $\mathrm{P}_{1,i,target}$ is a function of $\rho$,  the approximation developed in \eqref{policy I user 2 approximation} is still applicable, as explained in the following. With $\mathrm{P}_{1,i,target}=1-e^{-\frac{x\epsilon_{1,i}}{\rho}}$, the power allocation coefficient $\beta_i$ becomes $\beta_i^2=\frac{1-\frac{1}{x}}{1+\epsilon_{1,i}}$.  When $\rho$ approaches infinity, $\xi_m=\frac{x\epsilon_{1,i}}{\rho}$ is approaching zero, and the same conclusion can be made for $\frac{\epsilon_{2,m}}{\rho \beta_m^2 }$. As a result, the diversity order shown in \eqref{policy I user 2 approximation} is also applicable to the case with $\mathrm{P}_{1,i,target}=1-e^{-\frac{x\epsilon_{1,i}}{\rho}}$.    

\subsection{Power Allocation Policy II}
With this type of power allocation, the power allocation coefficients become functions of the instantaneous channel gains, and this fact  makes the evaluation of the outage probability very challenging, as explained in the following. First define
$y_{ii}=\left[\left(\mathbf{V}_2^H \mathbf{H}_{1}^H\mathbf{H}_{1}\mathbf{V}_2\right)^{-1}\right]_{i,i} $. As a result, the power allocation coefficient for user $2$ can be expressed as follows:
{\small\begin{align}\label{beta choice2}
\beta_{i}^2 =
\max\left\{0, \min\left\{\frac{y_{ii}\left(\frac{1}{y_{ii}}-\frac{\epsilon_{1,i}}{\rho}\right)}{(1+\epsilon_{1,i})},
\frac{x_i-\frac{\epsilon_{1,i}}{\rho}}{x_i(1+\epsilon_{1,i})}, \right\}\right\}.
\end{align}}
Even if we can reduce the expression  of $\beta$ to $\beta_{i}^2 =
\max\left\{0,  \frac{y_{ii}\left(\frac{1}{y_{ii}}-\frac{\epsilon_{1,i}}{\rho}\right)}{(1+\epsilon_{1,i})} \right\}$, a policy  conventionally used in \cite{Zhiguo_CRconoma}, the power allocation coefficient $\beta_i$ is still a function of $y_{ii}$, which means that the outage probability for user $2$ to detect $w_i$ can be written as follows:
\begin{align} \nonumber
\mathrm{P}_{2,i}^o=& \underset{y_{11},\cdots,y_{ii}}{\int\cdots\int} \sum^{i}_{m=1}
\left[1-e^{-\gamma_m}\left(\sum^{M-m}_{j=0}\frac{\gamma_m^j}{j!}\right)\right] \\ \nonumber &\times \prod^{m-1}_{n=1}\left[e^{-\gamma_n}\left(\sum^{M-n}_{j=0}\frac{\gamma_n^j}{j!}\right)\right]\\ \label{eqx} &\times f_{y_{11},\cdots, y_{ii}}(y_{11},\cdots, y_{ii})dy_{11}\cdots dy_{ii},
\end{align}
where the outage probability expression in \eqref{policy I user 2 approximation} is used, and $f_{y_{11},\cdots, y_{ii}}(y_{11},\cdots, y_{ii})$ is the joint pdf of $(y_{11},\cdots, y_{ii})$. Recall that $y_{ii}$ is the $i$-th element on the diagonal of the inverse Wishart matrix $\mathbf{W}^{-1}\triangleq\left(\mathbf{V}_2^H \mathbf{H}_{1}^H\mathbf{H}_{1}\mathbf{V}_2\right)^{-1}$. Note that the joint pdf can be obtained by calculating the marginal pdf of $\mathbf{W}^{-1}$ as follows: \cite{861781}
\begin{align}\label{correlations}
&f_{y_{11},\cdots, y_{ii}}(y_1,\cdots, y_i) \\ \nonumber =&\underset{y_{ij}, \forall i\neq j}{\int\cdots\int} \frac{\left(\det \mathbf{W}^{-1}\right)^{2N}}{\Gamma} e^{-\text{tr}\left(\mathbf{W}^{-1}\right)} dy_{12} \cdots dy_{N(N-1)}
\end{align}
where $\Gamma=\pi^{\frac{N(N-1)}{2}}\prod^{N}_{j=1}\Gamma(N-j+1)$.

Because of the correlation among $y_{ii}$ shown in \eqref{correlations} and the complicated form of the power allocation coefficients in \eqref{beta choice2}, a closed-form expression for the outage probability for user $2$ to decode $w_i$ cannot be found.   In the following, we will focus on the development of   upper and lower  bounds on the outage probability, which will be used for the analysis of the diversity gain achieved by the proposed MIMO-NOMA scheme at user $2$.

\subsubsection{Upper and lower bounds on the outage probability}\label{subsection bounds}
By using the definitions provided in Section \ref{subsection p1}, the outage probability for user $2$ to decode its own message  at the $i$-th layer can be expressed as follows:
\begin{align}\label{t1}
\mathrm{P}_{2,i}^o=& \sum^{i}_{m=1} \left(\mathrm{P}\left( \bar{E}_{m,1} \right)+\mathrm{P}\left( \bar{E}_{m,2} \right)\right).
\end{align}
In the following, we first focus on the development of an upper bound on the outage probability.
The   probability, $\mathrm{P}\left( \bar{E}_{m,1} \right)$, can be upper bounded as follows:
\begin{align}
\mathrm{P}\left( \bar{E}_{m,1} \right) &\leq \mathrm{P}\left(\log\left(1+\text{SINR}_{2,m'}\right)<R_{1,m}\right).
\end{align}
Similarly we can upper bound $\mathrm{P}\left( \bar{E}_{m,2} \right)$ as follows:
\begin{align}
\mathrm{P}\left( \bar{E}_{m,2} \right) &\leq \mathrm{P}\left(\log\left(1+\text{SNR}_{2,m}\right)<R_{2,m},\right.\\ \nonumber &\left.\log\left(1+\text{SINR}_{2,m'}\right)>R_{1,m} \right).
\end{align}

By using the SINR expression in \eqref{SINR}, the upper bound on $\mathrm{P}\left( \bar{E}_{m,1} \right)$ can be expressed as follows:
\begin{align}\label{indepdentx}
\mathrm{P}\left( \bar{E}_{m,1} \right)& \leq \underset{Q_1}{\underbrace{\mathrm{P}\left( z_m< x_m<  \frac{\frac{\epsilon_{1,m}}{\rho}}{\alpha^2_m - \beta_{m}^2\epsilon_{1,m}}  \right)}}  \\\nonumber &+  \underset{Q_2}{\underbrace{\mathrm{P}\left( z_m>x_m, x_m<  \frac{\frac{\epsilon_{1,m}}{\rho}}{\alpha^2_m - \beta_{m}^2\epsilon_{1,m}}  \right)}}  .
\end{align}
 The reason to have the two probabilities, $Q_1$ and $Q_2$, is that the power allocation coefficient $\beta_i$ has different forms depending on the relationship between $x_m$ and $z_m$.

By substituting the expression for $\beta_i$ when $z_m<x_m$, the factor $Q_1$ can be expressed as follows:
\begin{align}
Q_1 &= \mathrm{P}\left( z_m< x_m<  \frac{\frac{\epsilon_{1,m}}{\rho}}{1- \beta_{m}^2(1+\epsilon_{1,m})}  \right)
\\\nonumber &=
\mathrm{P}\left( z_m< x_m<  \frac{\frac{\epsilon_{1,m}}{\rho}}{1- \max\left\{0,  \frac{ \left(z_m-\frac{\epsilon_{1,m}}{\rho}\right)}{z_m(1+\epsilon_{1,m})} \right\}(1+\epsilon_{1,m})}  \right).
\end{align}
To simplify the outage probability, the  max function needs to be removed, and we   have the following:
\begin{align}\label{xx2}
Q_1   &=
\mathrm{P}\left( z_m< x_m<  \frac{\frac{\epsilon_{1,m}}{\rho}}{ \min\left\{1,  \frac{ \frac{\epsilon_{1,m}}{\rho} }{z_m} \right\}}  \right)\\\nonumber
&=
\mathrm{P}\left( z_m< x_m<   \max\left\{ \frac{\epsilon_{1,m}}{\rho} ,   z_m  \right\}\right)
\\\nonumber
&=
\mathrm{P}\left( z_m< x_m<   \frac{\epsilon_{1,m}}{\rho} \right)+
\mathrm{P}\left( \frac{\epsilon_{1,m}}{\rho} <z_m< x_m<   z_m  \right).
\end{align}
Note that the second probability in the above equation is zero. Since an upper bound is of interest, we   have
\begin{align}\label{t2}
Q_1   &\leq
\mathrm{P}\left( z_m<     \frac{\epsilon_{1,m}}{\rho} \right) \sim \frac{1}{\rho}.
\end{align}

The factor $Q_2$ can be calculated as follows:
\begin{align}\label{xx1}
Q_2&=\mathrm{P}\left( z_m>x_m, \right.\\\nonumber &\left. x_m<  \frac{\frac{\epsilon_{1,m}}{\rho}}{1-   \max\left\{0,  \frac{ \left(x_m-\frac{\epsilon_{1,m}}{\rho}\right)}{x_m(1+\epsilon_{1,m})} \right\}(1+\epsilon_{1,m})} \right)
\\\nonumber
&=\mathrm{P}\left( z_m>x_m, x_m<      \max\left\{\frac{\epsilon_{1,m}}{\rho} ,  x_m \right\}  \right).
\end{align}
By using the two possible choices of $x_m$, the above probability can be further upper bounded as follows:
\begin{align}\label{t3}
Q_2&=\mathrm{P}\left( z_m>x_m, x_m<       \frac{\epsilon_{1,m}}{\rho} , \frac{\epsilon_{1,m}}{\rho} > x_m  \right)
\\\nonumber &\leq \mathrm{P}\left(  x_m< \frac{\epsilon_{1,m}}{\rho}   \right) \sim \frac{1}{\rho^{M-m+1}},
\end{align}
where the first equation follows from the fact that  $\max\left\{\frac{\epsilon_{1,m}}{\rho} ,  x_m \right\}=x_m$ for  $ \frac{\epsilon_{1,m}}{\rho} < x_m $, a situation in which  user $2$ can decode $s_m$ for sure, i.e.,
$$\mathrm{P}\left( z_m>x_m, x_m<      \max\left\{\frac{\epsilon_{1,m}}{\rho} ,  x_m \right\}  \right)=0,$$  for   $ \frac{\epsilon_{1,m}}{\rho} < x_m $.

On the other hand, $\mathrm{P}\left( \bar{E}_{m,2} \right)$ can be calculated as follows:
\begin{align}\nonumber
&\mathrm{P}\left( \bar{E}_{m,2} \right) \leq
\underset{Q_3}{\underbrace{\mathrm{P}\left( z_m<x_m,   \frac{\frac{\epsilon_{1,m}}{\rho}}{\alpha^2_m - \beta_{m}^2\epsilon_{1,m}} <x_m< \frac{\epsilon_{2,m}}{\beta_m^2\rho} \right)}}\\
&+
\underset{Q_4}{\underbrace{\mathrm{P}\left(  z_m>x_m,  \frac{\frac{\epsilon_{1,m}}{\rho}}{\alpha^2_m - \beta_{m}^2\epsilon_{1,m}} <x_m< \frac{ \epsilon_{2,m}}{\beta_m^2\rho} \right)}}.
\end{align}
The factor $Q_3$ can be written as follows:
\begin{align}\nonumber
Q_3 &=
\mathrm{P}\left(    \frac{\frac{\epsilon_{1,m}}{\rho}}{1 - \max\left\{0,  \frac{ \left(z_m-\frac{\epsilon_{1,m}}{\rho}\right)}{z_m(1+\epsilon_{1,m})} \right\}(1+\epsilon_{1,m})} <x_m< \right.\\\nonumber &\left.\frac{\epsilon_{2,m}}{\max\left\{0,  \frac{ \left(z_m-\frac{\epsilon_{1,m}}{\rho}\right)}{z_m(1+\epsilon_{1,m})} \right\}\rho},z_m<x_m \right)\\
&=\label{ours}
\mathrm{P}\left( z_m<\frac{\epsilon_{1,m}}{\rho}, \frac{\epsilon_{1,m}}{\rho}  <x_m,z_m<x_m \right)\\\nonumber
&+
\mathrm{P}\left(  z_m>\frac{\epsilon_{1,m}}{\rho},  z_m <x_m<  \frac{\epsilon_{2,m}}{   \frac{ \left(z_m-\frac{\epsilon_{1,m}}{\rho}\right)}{z_m(1+\epsilon_{1,m})}  \rho},z_m<x_m \right)\\\nonumber &\leq
\mathrm{P}\left( z_m<\frac{\epsilon_{1,m}}{\rho}  \right) +
\mathrm{P}\left(    z_m  <  \frac{\epsilon_{2,m}}{   \frac{ \left(z_m-\frac{\epsilon_{1,m}}{\rho}\right)}{z_m(1+\epsilon_{1,m})}  \rho}  \right) .
\end{align}
It is interesting to observe that the second probability in the above equation can be rewritten as follows:
\begin{align}\label{x332}
&\mathrm{P}\left(    z_m  <  \frac{\epsilon_{2,m}}{   \frac{ \left(z_m-\frac{\epsilon_{1,m}}{\rho}\right)}{z_m(1+\epsilon_{1,m})}  \rho}  \right) \\\nonumber&=\mathrm{P}\left(    z_m <\frac{\epsilon_{1,m}+\epsilon_{2,m}+\epsilon_{1,m}\epsilon_{2,m}}{\rho} \right).
\end{align}
Therefore, the factor $Q_3$ can be upper bounded as follows:
\begin{align}\nonumber
Q_3  &\leq
\mathrm{P}\left( z_m<\frac{\epsilon_{1,m}}{\rho}  \right) +\mathrm{P}\left(    z_m <\frac{\epsilon_{1,m}+\epsilon_{2,m}+\epsilon_{1,m}\epsilon_{2,m}}{\rho} \right)\\ \label{t4}  &\sim\frac{1}{\rho}.
\end{align}
Furthermore, the factor $Q_4$ can be calculated as follows:
\begin{align}
Q_4=&
\mathrm{P}\left(   \frac{\frac{\epsilon_{1,m}}{\rho}}{1 - \max\left\{0,  \frac{ \left(x_m-\frac{\epsilon_{1,m}}{\rho}\right)}{x_m(1+\epsilon_{1,m})} \right\}(1+\epsilon_{1,m})} <\right.\\\nonumber &\left. x_m<\frac{ \epsilon_{2,m}}{\max\left\{0,  \frac{ \left(x_m-\frac{\epsilon_{1,m}}{\rho}\right)}{x_m(1+\epsilon_{1,m})} \right\}\rho} , z_m>x_m\right)\\\nonumber &\underset{(a)}{=}
\mathrm{P}\left(x_m<\frac{\epsilon_{1,m}}{\rho},    \frac{\epsilon_{1,m}}{\rho}  <  x_m, z_m>x_m\right)\\\nonumber &+
\mathrm{P}\left( x_m>\frac{\epsilon_{1,m}}{\rho},    x_m<\frac{ \epsilon_{2,m}}{   \frac{ \left(x_m-\frac{\epsilon_{1,m}}{\rho}\right)}{x_m(1+\epsilon_{1,m})}  \rho} , z_m>x_m\right)\\\nonumber &\leq
\mathrm{P}\left(x_m<\frac{\epsilon_{1,m}}{\rho}\right)+\mathrm{P}\left(      x_m<\frac{ \epsilon_{2,m}}{   \frac{ \left(x_m-\frac{\epsilon_{1,m}}{\rho}\right)}{x_m(1+\epsilon_{1,m})}  \rho}  \right),
\end{align}
where the step (a) is due to  the fact that, when $x_m>\frac{\epsilon_{1,m}}{\rho}$ and $z_m>x_m$,  user $2$ can always decode $s_m$,  since $\log(1+\text{SINR}_{2,m'})=R_{1,m}$.

According to \eqref{x332}, the factor $Q_4$ can be upper bounded as follows:
\begin{align}\nonumber
Q_4 &\leq
\mathrm{P}\left(x_m<\frac{\epsilon_{1,m}}{\rho}\right)+\mathrm{P}\left(    x_m <\frac{\epsilon_{1,m}+\epsilon_{2,m}+\epsilon_{1,m}\epsilon_{2,m}}{\rho} \right)\\ &\sim \frac{1}{\rho^{M-m+1}}. \label{t5}
\end{align}
By combining \eqref{t1}, \eqref{t2}, \eqref{t3}, \eqref{t4} and \eqref{t5}, we can conclude that a lower bound on the diversity gain  at user $2$, obtained from the upper bound on the outage probability, is $1$.

A lower bound on the outage probability can obtained as follows:
\begin{align}
\mathrm{P}_{2,i}^o\geq&  \mathrm{P}\left( \bar{E}_{1,2} \right) \geq Q_3,
\end{align}
by focusing the case of $m=1$.
Following \eqref{ours}, the factor $Q_3$ with $m=1$ can be calculated as follows:
\begin{align}
Q_3   &\geq \mathrm{P}\left( z_1<\frac{\epsilon_{1,1}}{\rho}, \frac{\epsilon_{1,1}}{\rho}  <x_1  \right) \\\nonumber &=
\mathrm{P}\left( z_1<\frac{\epsilon_{1,1}}{\rho}  \right)-\mathrm{P}\left( z_1<\frac{\epsilon_{1,1}}{\rho}, x_1 <\frac{\epsilon_{1,1}}{\rho} \right) \\\nonumber &=
\mathrm{P}\left( z_1<\frac{\epsilon_{1,1}}{\rho}  \right)\left(1- \mathrm{P}\left( x_1 <\frac{\epsilon_{1,1}}{\rho} \right)\right)\sim\frac{1}{\rho},
\end{align}
since $z_1$ is independent of $x_1$, $\mathrm{P}\left( z_1<\frac{\epsilon_{1,1}}{\rho}  \right)\sim \frac{1}{\rho}$ and $\mathrm{P}\left( x_1 <\frac{\epsilon_{1,1}}{\rho} \right)\sim\frac{1}{\rho^M}$.

Since both upper and lower bounds converge, we can conclude that the diversity gain for user $2$ to decode $w_i$   is one, when power allocation policy II is used.

 {\it Remark 4:} Note that the diversity gain obtained above is the same as that at user $1$, when power allocation policy II is used. This is consistent with the conclusion made in \cite{Zhiguo_CRconoma}, where the diversity gain at the user with stronger channel conditions is determined by the channel conditions of its partner, when the cognitive radio inspired power allocation policy is applied.

\section{Numerical Studies}\label{section simulation}
In this section, the performance of the proposed MIMO-NOMA scheme is evaluated by using simulation results. We will first compare the proposed scheme with some existing MIMO-NOMA and MIMO-OMA schemes. Then, additional  simulation results are provided to demonstrate the impact of different choices of the system parameters, where analytical results developed in the paper will also be verified.

\subsection{Comparison to Benchmark Schemes}
To simplify the simulation comparison,   power allocation policy I is used in this subsection. The targeted data rates for two users are set as $R_{1,i}=1$ bit per channel user (BPCU) and $R_{2,i}=4$ BPCU, for all $1\leq i\leq N$, respectively. The targeted outage probabilities for the two users are set as $\mathrm{P}_{1,i,target}=1-e^{-\frac{2\epsilon_{1,i}}{\rho}}$, and $\mathrm{P}_{2,i,target}=1-e^{-\frac{2\epsilon_{2,i}}{\rho}}$, respectively. Using these targeted outage probabilities and the step in \eqref{betax1}, the power allocation coefficients can be obtained. Note that these chosen $\mathrm{P}_{k,i,target}$ are still within the range defined in \eqref{range1}. Since  the use of power allocation policy I guarantees the QoS requirements at user $1$,  we will focus on the outage performance at user $2$   in this subsection.

We first compare the proposed  scheme to those MIMO-NOMA schemes developed  in \cite{Zhiguo_mimoconoma} and \cite{Dingicc16}, which are termed ZF-NOMA and SA-NOMA, respectively. Since both schemes were proposed for scenarios with different system parameters, they need be tailored to the scenario addressed in this paper as explained in the following. Recall that ZF-NOMA proposed in  \cite{Zhiguo_mimoconoma} requires $N\geq M$, and SA-NOMA proposed in \cite{Dingicc16} requires $N>\frac{M}{2}$. In order to ensure that both schemes are applicable, we focus on a scenario with $N=M$, but it is important to point out that the scheme proposed in this paper is applicable to a scenario with a small $N$.

For ZF-NOMA, the precoding matrix is set as an identity matrix, i.e., $\mathbf{P}=\mathbf{I}_M$, and both users use zero forcing for detection. The SINR for user $2$ to decode the message intended to user $1$ at the $i$-th layer can be written as $\text{SINR}_{ZF,i}=\frac{\frac{\alpha_i^2}{\left[\left( \mathbf{H}_{2}^H\mathbf{H}_{2}\right)^{-1}\right]_{i,i}}}{\frac{\beta_i^2}{\left[\left( \mathbf{H}_{2}^H\mathbf{H}_{2}\right)^{-1}\right]_{i,i}}+\frac{1}{\rho}}$. If user $2$ can decode its partner's message successively, it can decode its own with the following SNR: $\text{SNR}_{ZF,i}=\frac{\rho\beta_i^2}{\left[\left( \mathbf{H}_{2}^H\mathbf{H}_{2}\right)^{-1}\right]_{i,i}}$. In order to have a fair comparison, for ZF-NOMA, the effective channel gains are ordered, i.e., $\frac{1}{\left[\left( \mathbf{H}_{2}^H\mathbf{H}_{2}\right)^{-1}\right]_{1,1}}\geq \cdots \geq \frac{1}{\left[\left( \mathbf{H}_{2}^H\mathbf{H}_{2}\right)^{-1}\right]_{M,M}}$.

It is interesting to point out that for the case of $M=N$,   SA-NOMA achieves the same performance as ZF-NOMA, as shown in the following.  For SA-NOMA,   signal alignment is used, where  the two users' detection matrices, $\mathbf{U}_1$ and $\mathbf{U}_2$, are obtained from the equation $\begin{bmatrix}\mathbf{H}_1^H &-\mathbf{H}_2^H \end{bmatrix}\begin{bmatrix} \mathbf{U}_1 &\mathbf{U}_2 \end{bmatrix}^H=\mathbf{0}_{M\times M}$, where both detection matrices are $N\times N$. As a result, the users' effective channel matrices become the same, i.e., $\mathbf{U}_1\mathbf{H}_1=\mathbf{U}_2\mathbf{H}_2$. Therefore,  the SINR for user $2$ to decode the message intended for user $1$ at the $i$-th layer can be written as $\text{SINR}_{SA,i}= \frac{\alpha_i^2}{\beta_{i}^2 +\frac{1}{\rho} \left[ (\mathbf{U}_2\mathbf{H}_2)^{-1}\mathbf{U}_2\mathbf{U}_2^H(\mathbf{U}_2\mathbf{H}_2)^{-H}\right]_{i,i}} =\text{SINR}_{ZF,i}$, where both $\mathbf{H}_i$ and $\mathbf{U}_i$ are assumed to be invertible. Similarly, provided that user $2$ can decode its parter's message, it can decode its own with the following SNR: $\text{SNR}_{SA,i}= \frac{\rho\beta_i^2}{\left[ (\mathbf{U}_2\mathbf{H}_2)^{-1}\mathbf{U}_2\mathbf{U}_2^H(\mathbf{U}_2\mathbf{H}_2)^{-H}\right]_{i,i}} =\text{SNR}_{ZF,i}$. Therefore, the two schemes achieve the same outage performance in the addressed scenario with $M=N$.

\begin{figure}[!htp]
\begin{center} \includegraphics[width=0.47\textwidth]{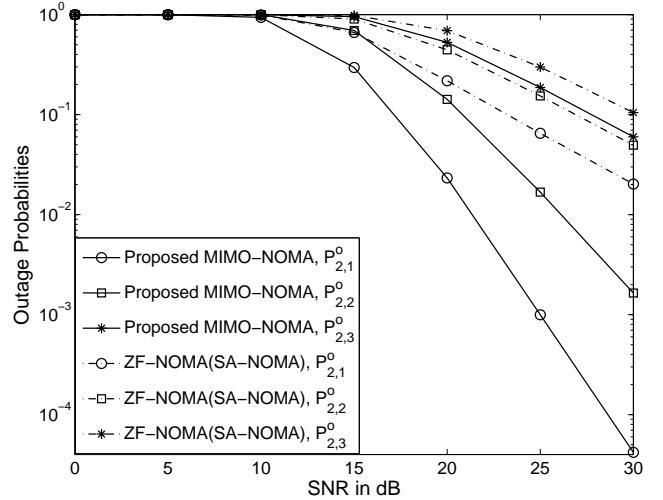}
\end{center}
\vspace*{-2mm} \caption{Comparison to the existing MIMO-NOMA schemes. $M=N=3$.   $\mathrm{P}_{1,i,target}=1-e^{-\frac{2\epsilon_{1,i}}{\rho}}$, and $\mathrm{P}_{2,i,target}=1-e^{-\frac{2\epsilon_{2,i}}{\rho}}$. $R_{1,i}=1$ BPCU and $R_{2,i}=4$ BPCU, $\forall~1\leq i\leq N$.  }\label{fig com 1}
\end{figure}

In Fig. \ref{fig com 1}, the performance of these MIMO-NOMA schemes is shown as a function of the transmit SNR. As can be seen from the figure, for all MIMO-NOMA schemes considered, the outage performance  at the $i$-th layer  is better than that at the $j$-th layer, for $i<j$, which can be explained as follows. For the proposed scheme,  the effective channel gain at the $i$-th layer, $[\mathbf{R}_2^H]_{i,i}^2$, is statistically stronger than that at the $j$-th layer, since $[\mathbf{R}_2^H]_{i,i}^2$ is chi-square distributed with $2(M-i+1)$ degrees of freedom. For the two existing MIMO-NOMA schemes, we have ordered the effective channel gains as $\frac{1}{\left[\left( \mathbf{H}_{2}^H\mathbf{H}_{2}\right)^{-1}\right]_{1,1}}\geq \cdots \geq \frac{1}{\left[\left( \mathbf{H}_{2}^H\mathbf{H}_{2}\right)^{-1}\right]_{M,M}}$. Furthermore, it is important to observe that at all layers, the proposed scheme outperforms the existing MIMO-NOMA schemes. Particularly, the figure demonstrates  that for the proposed scheme, the slope of the outage probability curves is changing, which means  change of the diversity gains. On the other hand, all the outage probability curves for the existing schemes have the same slope, which is mainly due to the correlation among  the effective channel gains, $\frac{1}{\left[\left( \mathbf{H}_{2}^H\mathbf{H}_{2}\right)^{-1}\right]_{i,i}}$.

OMA is another important benchmark scheme. Recall that in this paper, user $1$ is viewed  as a primary user in a conventional cognitive radio network. If OMA is used, the bandwidth resource allocated to user $1$ cannot be reused. The use of NOMA means that user $2$, which can be viewed as a secondary user, is admitted into  the bandwidth occupied by user $1$.  Because the proposed power allocation policies can ensure that user $1$'s QoS requirements are met, whatever can be transmitted to user $2$, such as $R_{2,i}(1-\mathrm{P}_{2,i}^o)$, will be a net performance  gain over OMA. Or in other words, the benefit of the proposed MIMO-NOMA scheme over OMA is clear if we ask user $2$, a user with stronger channel conditions, to be admitted into the bandwidth allocated to user $1$.

In the following, we consider a comparison which is more difficult for NOMA. Particularly, consider that there are two time slots (or frequency-channels/ spreading-codes). For OMA, time slot $i$ is allocated to user $i$. For NOMA, the two users are served at the same time. Comparing NOMA to this type of OMA is challenging, since user $1$, a user with weaker channel conditions, is  admitted  into the time slot allocated to user $2$ and user $2$ cannot enjoy interference free communications experienced in the OMA case. As a result, the performance gain of NOMA over OMA becomes less  obvious.

\begin{figure}[!htp]
\begin{center} \subfigure[ $M=N=3$]{\includegraphics[width=0.47\textwidth]{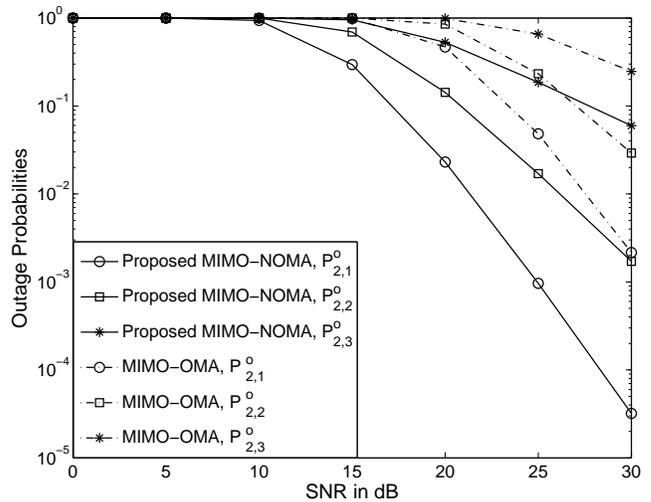}}
\subfigure[$M=6$ and $N=3$]{
\includegraphics[width=0.47\textwidth  ]{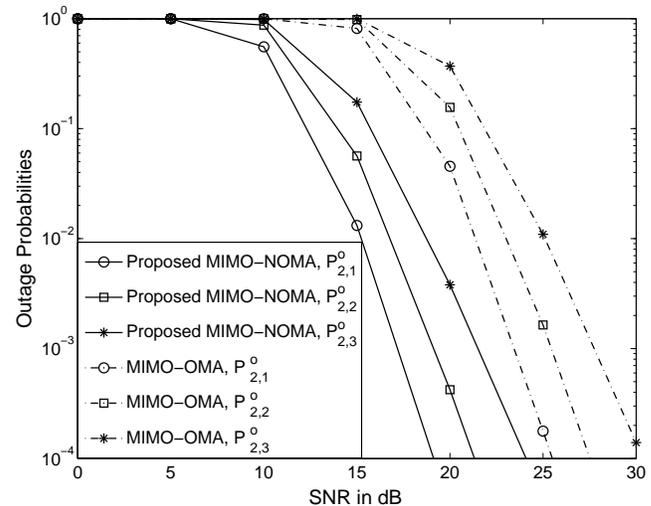}}
\end{center}
\vspace*{-3mm} \caption{Comparison to    MIMO-OMA.     $\mathrm{P}_{1,i,target}=1-e^{-\frac{2\epsilon_{1,i}}{\rho}}$, and $\mathrm{P}_{2,i,target}=1-e^{-\frac{2\epsilon_{2,i}}{\rho}}$. $R_{1,i}=1$ BPCU and $R_{2,i}=4$ BPCU, $\forall~1\leq i\leq N$.  }\label{fig com 2}
\end{figure}

In Fig. \ref{fig com 2}, the performance of the proposed MIMO-NOMA scheme is compared to the MIMO-OMA scheme described above. Precoding for the considered  MIMO-OMA scheme is designed   by using the same QR approach as discussed in Section~\ref{section system model}. Particularly, during the second time slot, user $2$ is served and the precoding matrix is designed according to the QR decomposition of $\mathbf{H}_2$, which means that the data rate for user $2$ to decode its message at the $i$-th layer is $\frac{1}{2}\log(1+\rho[\mathbf{R}_{2}^H]^2_{i,i})$, where the factor $\frac{1}{2}$ is due to the use of OMA. As can be seen from the two sub-figures in Fig.\ref{fig com 2}, the proposed MIMO-NOMA scheme can achieve better outage performance compared to MIMO-OMA, and the performance gap between the two schemes can be increased by introducing more antennas  at the base station, i.e., increasing $M$. An interesting observation is that the slope of the outage probability curves for MIMO-NOMA is the same as that for MIMO-OMA, which means that the diversity gains achieved by the two schemes are the same. This phenomenon  is expected since for both schemes, the outage probabilities are determined by the same effective channel gain, $[\mathbf{R}_{2}^H]^2_{i,i}$.

\subsection{Impact of Different  System Parameters on Users' Outage Performance}
\begin{figure}[!htp]
\begin{center} \includegraphics[width=0.47\textwidth]{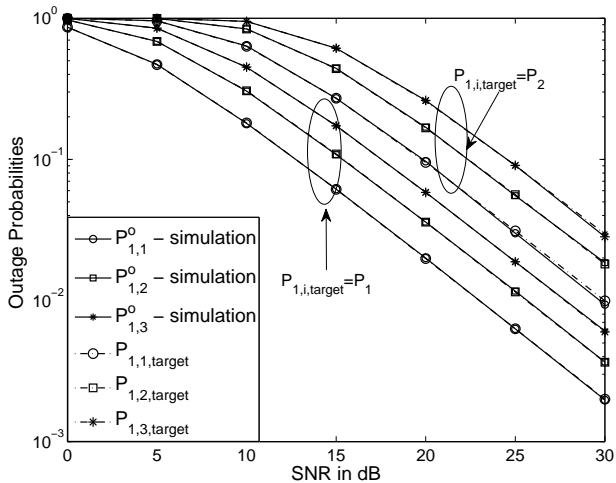}
\end{center}
\vspace*{-3mm} \caption{Outage performance at user $1$ with power allocation policy I.  $M=N=3$.   $\mathrm{P}_1=1-e^{-\frac{2\epsilon_{1,i}}{\rho}}$, and $\mathrm{P}_2=1-e^{-\frac{10\epsilon_{2,i}}{\rho}}$. $R_{1,1}=1$ BPCU, $R_{1,2}=1.5$ BPCU, and $R_{1,2}=2$ BPCU.   }\label{fig u1 1}
\end{figure}

\begin{figure}[!htp]
\begin{center} \includegraphics[width=0.47\textwidth]{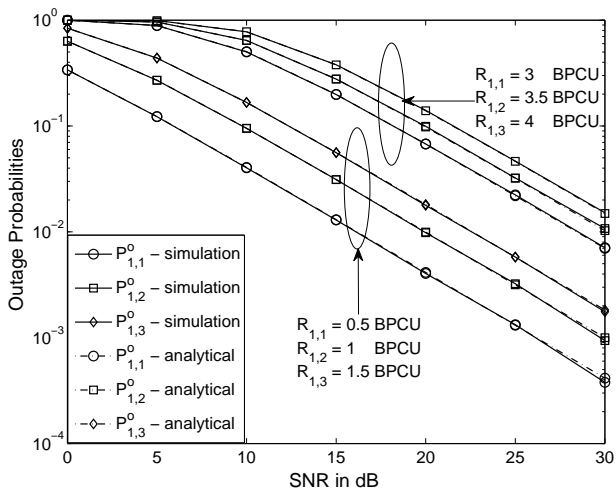}
\end{center}
\vspace*{-3mm} \caption{Outage performance at user $1$ with power allocation policy II.  $M=N=3$.  The analytical results are based on \eqref{outage user 1}.  }\label{fig u1 2}
\end{figure}

In Figs. \ref{fig u1 1} and \ref{fig u1 2}, user $1$'s outage performance achieved by  the proposed MIMO-NOMA scheme is shown with different choices of the targeted data rates. Particularly, when  power allocation policy I is used, Fig. \ref{fig u1 1} shows that the outage probability curves achieved by the proposed MIMO-NOMA transmission scheme match perfectly with those for the targeted outage probability, which demonstrates that the proposed transmission scheme can  strictly guarantee the QoS requirements at user $1$ in the long term. When power allocation policy II is used, Fig. \ref{fig u1 2} demonstrates that the simulation results match perfectly with the analytical results developed in  \eqref{outage user 1}. Therefore, the use of power allocation policy II guarantees that the outage probability at user $1$ is $\left(1-e^{-\frac{\epsilon_{1,i}}{\rho}}\right)$, which is equivalent to the outage performance for the case in which all the power is allocated  to user $1$.

\begin{figure}[!htp]
\begin{center} \includegraphics[width=0.47\textwidth]{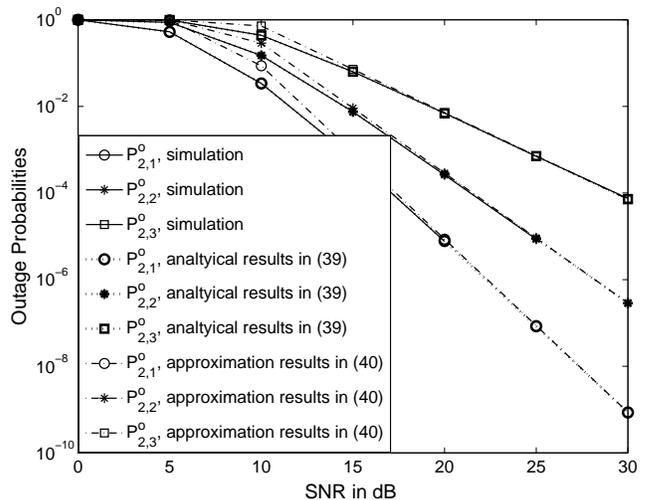}
\end{center}
\vspace*{-3mm} \caption{Outage performance at user $2$ with power allocation policy I. $M=N=3$, $\mathrm{P}_1=1-e^{-\frac{2\epsilon_{1,i}}{\rho}}$, $\mathrm{P}_2=1-e^{-\frac{10\epsilon_{2,i}}{\rho}}$, $R_{1,i}=1$ BPCU, and $R_{2,i}=2$ BPCU.   The analytical results are based on \eqref{policy I user 2} and \eqref{policy I user 2 approximation}.  }\label{fig u2 1}
\end{figure}

In Fig. \ref{fig u2 1}, the outage probabilities at user $2$ are shown as  functions of the transmit SNR, when power allocation policy I is used. As can be observed from the figure, the curves for the simulation results match perfectly with the ones for the analytical result developed in \eqref{policy I user 2}, which demonstrates the accuracy of this exact expression for the outage probability. The curves for the approximation results developed in \eqref{policy I user 2 approximation} match the simulation curves  only at high SNR, which is due to the fact that this approximation is obtained with the high SNR assumption.  Another important observation from this figure is that the slope of the  outage probability  curve for $\mathrm{P}_{2,i}^o$ is larger than that of $\mathrm{P}_{2,j}^o$, for $i<j$. This is because a larger diversity gain can be obtained at the $i$-th layer, compared to that at the $j$-th layer, as discussed in Remark $3$ in Section \ref{subsection p1}.

\begin{figure}[!htp]
\begin{center} \includegraphics[width=0.47\textwidth]{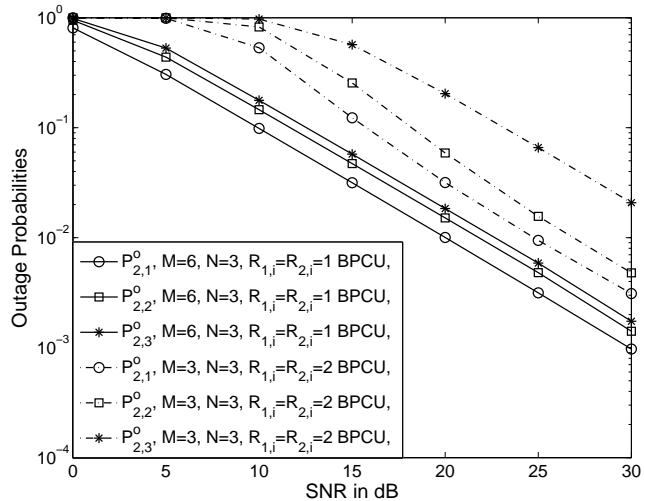}
\end{center}
\vspace*{-3mm} \caption{Outage performance at user $2$ with power allocation policy II.  $M=N=3$.    }\label{fig u2 2}
\end{figure}

\begin{figure}[!htp]
\begin{center} \includegraphics[width=0.47\textwidth]{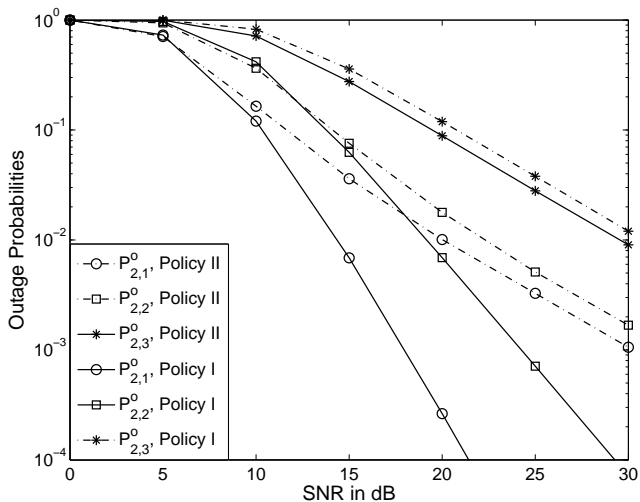}
\end{center}
\vspace*{-3mm} \caption{Comparison between the two power allocation policies.   $M=N=3$, $R_{1,i}=1$ BPCU,   $R_{2,i}=2$ BPCU, $\mathrm{P}_1=1-e^{-\frac{2\epsilon_{1,i}}{\rho}}$ and $\mathrm{P}_2=1-e^{-\frac{10\epsilon_{2,i}}{\rho}}$.    }\label{fig u12}
\end{figure}

Fig. \ref{fig u2 2} demonstrates the outage performance at user $2$ when power allocation policy II is used. As shown in the figure, the outage performance of user $2$ can be improved by increasing the number of antennas at the base station or decreasing the targeted data rates.  An important observation from this figure is that at high SNR, the slope of all the curves is the same, which means that the same diversity gain is achieved at all layers, regardless of the choice of the number of antennas at the base station. This confirms the analytical results developed in Section \ref{subsection bounds}, in which it is shown that the diversity gain is one for all layers. In Fig. \ref{fig u12}, the outage performance experienced by user $2$ with different power allocation policies is compared. Particularly, this figure shows that the use of power allocation policy I is preferable for user $2$, since better outage performance can be achieved. However, it is important to point out that the use of power allocation policy II can meet the QoS requirement of user $1$ instantaneously as shown in \eqref{policy II}, whereas power allocation policy I can only meet the long term QoS requirement.

%

\section{Conclusions}
In this paper, we have  considered a MIMO-NOMA downlink transmission scenario, where a new precoding and power allocation strategy was proposed to ensure that the potential of NOMA can be realized even if the participating users' channel conditions are similar. Particularly, the precoding matrix has been designed to degrade user $1$'s effective channel gains while improving the signal strength at user $2$. Two types of power allocation policies have been  developed to meet user $1$'s QoS requirement in a long and short term, respectively. Analytical and numerical results have also been  provided to demonstrate the advantages and disadvantages of the two power allocation policies. The outage performance at user $2$ has been  analyzed by using some bounding techniques, whereas an important future direction is to find a closed-form expression for this outage probability by applying the order statistics of the diagonal elements of an inverse Wishart matrix.

%
%
 \bibliographystyle{IEEEtran}
\bibliography{IEEEfull,trasfer}

  \end{document}